\documentclass[journal ]{new-aiaa}
\usepackage[utf8]{inputenc}
\usepackage{textcomp}
\usepackage{graphicx}
\usepackage{subcaption}
\usepackage{amsmath}
\usepackage{multicol}
\usepackage{multirow}
\usepackage{float}
\usepackage[version=4]{mhchem}
\usepackage{siunitx}
\usepackage{longtable,tabularx}
\usepackage{xcolor}

\usepackage{scalerel,stackengine}
\stackMath
\newcommand\reallywidehat[1]{%
\savestack{\tmpbox}{\stretchto{%
  \scaleto{%
    \scalerel*[\widthof{\ensuremath{#1}}]{\kern.1pt\mathchar"0362\kern.1pt}%
    {\rule{0ex}{\textheight}}
  }{\textheight}%
}{2.4ex}}%
\stackon[-6.9pt]{#1}{\tmpbox}%
}

\providecommand{\filo}[1]{\widetilde{#1}}
\providecommand{\filh}[1]{\widehat{#1}}
\providecommand{\brak}[1]{\left(#1\right)}

\title{Near-Field Wall-Modeled Large-Eddy Simulation of the NASA X-59 Low-Boom Flight Demonstrator}

\author{
  Emily Williams\footnote{Department of Energy Computational Science Graduate Fellow, Department of Aeronautics and Astronautics, AIAA Student Member.},
  Gonzalo Arranz\footnote{Postdoctoral Associate, Department of Aeronautics and Astronautics, AIAA Member.}, and
  Adri\'an Lozano-Dur\'an \footnote{Boeing Assistant Professor, Department of Aeronautics and Astronautics, AIAA Senior Member.}}
\affil{Massachusetts Institute of Technology, Cambridge, MA, 02142} 
\begin{document}

\maketitle

\begin{abstract}
Wall-modeled large-eddy simulation (WMLES) is utilized to analyze the
experimental aircraft X-59 Quiet SuperSonic Technology (QueSST)
developed by Lockheed Martin at Skunk Works for NASA's Low-Boom Flight
Demonstrator project. The simulations utilize the charLES solver and
aim to assess the ability of WMLES to predict near-field noise levels
under cruise conditions, considering various subgrid-scale (SGS)
models and grid resolutions.  The results are compared with previous
numerical studies based on the Reynolds-averaged Navier-Stokes (RANS) equations. Our findings demonstrate that WMLES produces
near-field pressure predictions that are similar to those of RANS
simulations at a comparable computational cost.  Some mild
discrepancies are observed between the WMLES and RANS predictions
downstream the aircraft.  These differences persist for finest grid
refinement considered, suggesting that they might be attributed to  underresolved interactions of shock waves and expansions waves at the
trailing edge.
\end{abstract}

\section*{Nomenclature}
{\renewcommand\arraystretch{1.0}
\noindent\begin{longtable*}{@{}l @{\quad=\quad} l@{}}
$\alpha_{ij}$ & resolved velocity gradient tensor \\
$C_{\mathrm{dsm}}$ & constant for dynamic Smagorinsky model \\
$C_{\mathrm{vre}}$ & constant for Vreman model \\
$C_p$ & specific heat at constant pressure \\
$C_v$ & specific heat at constant volume \\
$\delta_{ij}$ & Kronecker delta \\
$\gamma$ & ratio of specific heats \\
$h_{\mathrm{wm}}$ & wall-normal matching location for the wall model \\
$\kappa$ & K\'arm\'an constant \\
$L_{\mathrm{BODY}}$ & streamwise body-length of the X-59 model \\
$M$ & Mach number \\
$\mu$ & dynamic viscosity \\
$\mu_M$ & Mach angle \\
$\mu_{t,\mathrm{wm}}$ & wall-model dynamic eddy viscosity \\
$Pr$ & Prandtl number \\
$Pr_{t}$ & turbulent Prandtl number \\
$Pr_{t,\mathrm{wm}}$ & turbulent Prandtl number for the wall model \\
$p$ & pressure \\
$p_\infty$ & farfield pressure \\
$p_s$ & pressure measured by the probe \\
$q_j$ & heat flux vector \\
$R$ & gas constant \\
$Re$ & Reynolds number \\
$\rho$ & density \\
$\rho_\infty$ & farfield density \\
$S_{ij}$ & rate-of-strain tensor \\
$\sigma_{ij}$ & shear-stress tensor \\
$\tau_{ij}$ & subgrid-scale tensor \\
$\tau_{w,\mathrm{wm}}$ & wall shear stress for the wall model \\
$T$ & temperature \\
$T_{\mathrm{wm}}$ & wall-model temperature \\
$u_i$ & velocity component \\
$u_{\mathrm{wm}}$ & magnitude of the wall-parallel velocity for the wall model \\
$U_\infty$ & farfield streamwise velocity \\
$\varepsilon_p$ & relative difference in the pressure signature between the RANS and WMLES \\
$x_i$ & spatial coordinate \\
$X_N$ & transformed $x$-coordinate \\
$y_n$ & wall-normal direction for wall model \\
$\nu$ & kinematic viscosity \\
$\nu_t$ & eddy viscosity \\
CFD & computational fluid dynamics \\
DLR & German Aerospace Center \\
EQWM & equilibrium wall-stress model \\
ECS & environmental control system \\
LES & large-eddy simulation \\
M & Mach number based on the freestream\\
SA & Spalart-Allmaras \\
RANS & Reynolds-averaged Navier--Stokes equations \\
SGS & subgrid scale \\
WMLES & wall-modeled large-eddy simulation
\end{longtable*}}

\section{Introduction}

The use of computational fluid dynamics (CFD) for external aerodynamic
applications with increasing functionality and performance has greatly
improved our understanding and predictive capabilities of complex
flows~\cite{Mani2023}. However, applications to the modern aerospace
industry often entail the presence of multi-scale, complex flow
physics, which have since exposed limitations in even the most
advanced CFD solvers~\cite{Slotnick2014}. One outstanding challenge in
CFD is the prediction of quantities of interest in high-speed
vehicles.  In this work, we investigate the capability of wall-modeled
large-eddy simulation (WMLES) to predict the near-field pressure
signature of the X-59 Quiet SuperSonic Technology (QueSST).
 
The primary objective of NASA's Low-Boom Flight Demonstrator project
is to assess the viability of supersonic flight over land while
minimizing noise levels. In this regard, the X-59 QueSST experimental
aircraft, developed by Lockheed Martin at Skunk Works for NASA's
project, serves as a practical platform for validating high-speed
aircraft CFD~\cite{Park2016, Park2019,
  Carter2020}. Previous research efforts, documented in the American
Institute of Aeronautics and Astronautics (AIAA) Sonic Boom Prediction
Workshop series, have encompassed near-field CFD models and subsequent
atmospheric propagation to analyze noise signatures specific to the
X-59 QueSST~\cite{park2022lowboom}. The workshop involved utilizing
various CFD solvers to simulate near-field signatures, which were then
propagated to the ground for noise level calculations.  Currently, CFD
studies of the X-59 QueSST mostly rely on closure models for the
Reynolds-averaged Navier-Stokes (RANS) equations~\cite{Casey2000,
  Roy2006}, along with other variations~\cite{Spalart1997,
  Spalart2009}.  Here, we use as a reference the near-field RANS
simulations from \citet{kirz2022dlrtau}. The RANS cases were performed
with the German Aerospace Center (DLR) TAU code, which is based on an
unstructured finite-volume approach for solving the RANS equations on
hybrid grids \cite{schwamborn2006thedlr, kirz2022dlrtau}. An improved
second-order accurate advection upstream splitting method (AUSM)
upwind scheme was applied for the spatial discretization of the
convective fluxes, and an implicit lower upper symmetric Gauss Seidel
scheme is used for time stepping. The AUSM scheme is a flux splitting
scheme with an aim at removing numerical dissipation on a contact
discontinuity for accurate and robust resolution for shocks
\cite{wada1997anaccurate}. The gradients were computed using a Green
Gauss approach. The Spalart-Allmaras (SA)-negative turbulence model
was used.  For a comprehensive overview of the RANS simulations, refer
to the work by \citet{park2022lowboom}.

Current RANS methodologies have demonstrated high performance in
predicting shock waves far from their interactions with boundary
layers~\cite{Roy2006}. Yet, there is still no practical RANS model
that can accurately address the wide range of flow regimes relevant to
aerodynamic vehicles. These regimes include separated flows,
afterbodies, mean-flow three-dimensionality, shock waves, aerodynamic
noise, fine-scale mixing, laminar-to-turbulent transition, and
more. RANS predictions in these scenarios often lack consistency and
reliability, particularly when dealing with geometries and conditions
representative of the flight envelope of
airplanes~\cite{Slotnick2014}. One notable example is the inadequate
prediction of the onset and extent of three-dimensional separated flow
in wing-fuselage junctures, where RANS-based approaches have exhibited
poor performance~\citep{Rumsey2018, Lozano2022}. Furthermore, CFD
experiments with aircraft at high angles of attack have revealed that
RANS-based solvers struggle to accurately predict maximum lift, the
corresponding angle of attack, and the physical mechanisms underlying
stall. While algebraic RANS closures, such as the Baldwin–Lomax
model~\citep{Baldwin1978}, offer reasonable predictions at high Mach
number boundary layer flows~\citep{Rumsey2010}, they are still
deficient in capturing separated flows~\citep{Gaitonde1993}.

Recently, large-eddy simulation (LES) has gained momentum as a tool
for aerospace applications that might offer a higher degree of
generality compared to RANS approaches~\cite{Bose2018}.  In LES, the
large eddies containing most of the energy are directly resolved,
while the dissipative effect of the small scales is modeled through a
subgrid-scale (SGS) model.  By incorporating a near-wall flow model
that resolves only the large-scale motions in the outer region of the
boundary layer, WMLES reduces the grid-point
requirements, scaling at most linearly with increasing Reynolds
number~\cite{Choi2012, Yang2021}.  Some representative studies of
WMLES at high-speed conditions include shock/boundary-layer
interactions~\cite{Bermejo2014, Zebiri2020, Mettu2019}, compressible
channels~\cite{Bocquet2012, Iyer2019}, the shock wave boundary layer
interaction over compression ramps~\citep{Dawson2013, Edwards2008},
flow past a single fin~\citep{Fang2015}, flow past two parallel
fins~\citep{Fu2020}, turbulent boundary layers~\citep{Li2013,
  Helm2022} and turbulent channel flows~\citep{Yang2018}, transitional
flows~\cite{Fu2018}, the  wall-mounted hump and axisymmetric compression
ramp~\cite{Iyer2016, Iyer2018}, and transonic
airfoils~\cite{Fukushima2018}.  Although most SGS/wall models are
rooted in aero/thermal equilibrium assumptions, which are only valid
for incompressible flows, the general consensus is that WMLES is able
to capture key flow physics.  Nonetheless, the use of WMLES for the
prediction of realistic supersonic aircraft has been explored to a
lesser extent, which is the goal of this work. 

Our long-term objective is to develop a SGS model for
WMLES that provides accurate
predictions across a wide range of flow regimes, including subsonic
and supersonic flows~\cite{Lozano_brief_2020, Ling2022, Arranz2023,
  Lozano2023}. The results from the present study will inform the
advances required to extend our SGS modeling approach in the presence
of shock waves.

This paper is organized as follows. The methodology of WMLES is
presented in section~\ref{sec:meth}, which contains details of the
numerical solver, and the physical and
computational setups are discussed in section ~\ref{sec:setup}. The numerical results for the quantities of
interest are presented in section~\ref{sec:results}. Finally,
conclusions and future work are offered in
section~\ref{sec:conclusion}.

\section{Methodology}
\label{sec:meth}

\subsection{SGS Models}

The performance of WMLES is investigated for two SGS models. Both
models adopt the eddy viscosity form
\begin{align}
    \tau_{ij} = -2 \nu_t \left( \widetilde{S}_{ij} - \frac{1}{3}\widetilde{S}_{kk}\delta_{ij}\right) + \frac{1}{3}\tau_{kk}\delta_{ij},
\end{align}
where repeated indices imply summation, $\tau_{ij}$ the SGS tensor,
$\nu_t$ is the eddy viscosity, $S_{ij}$ is the rate-of-strain,
$\tau_{kk}\delta_{ij} / 3$ is the isotropic component of the SGS
tensor with Kronecker delta $\delta_{ij}$, and $\widetilde{\phi} =
\overline{ \rho \phi}/\overline{\rho}$ is the Favre average with
$\overline{(\cdot)}$ denoting the grid filter. 

The first SGS model considered is the dynamic Smagorinsky model
(labeled as DSM)~\cite{Moin1991}, where the eddy viscosity is given by
$\nu_t = C_{\mathrm{dsm}} |\widetilde{S}|$.  The constant
$C_{\mathrm{dsm}}$ is dynamically
%
%
computed as \citep{Lilly1992}
\begin{align}
    C_{\mathrm{dsm}}  = \frac{1}{2}\frac{\left\langle \mathcal{L}_{ij} \mathcal{M}_{ij} \right\rangle}
    {\langle \mathcal{M}_{ij}^2 \rangle}
\end{align}
where
\begin{align*}
    \mathcal{L}_{ij} &= \overline{\rho} \brak{ \filh{\filo{u}_i \filo{u}_j} - \filh{\filo{u}}_i \filh{\filo{u}}_j}, &
    \mathcal{M}_{ij} &= \brak{\filh{\Delta}/\Delta}^2
    | \filh{\filo{S}} | \brak{ \filh{\filo{S}}_{ij} - \frac{1}{3}\filh{\filo{S}}_{kk} } - 
    \reallywidehat{ | \filo{S} |\brak{ {\filo{S}}_{ij} - \frac{1}{3}{\filo{S}}_{kk} }},
\end{align*}
with  $\widehat{(\cdot)}$ being the test filter with filter width
$\filh{\Delta}$ larger than that of the grid filter $\Delta$, and $|
\filo{S} |^2 = 2 \filo{S}_{ij}\filo{S}_{ij}$. 

The second SGS model considered is the Vreman
model~\cite{vreman2004aneddy} (labeled as Vreman). The eddy viscosity
is given by
\begin{align}
    \nu_t = C_{\mathrm{vre}} \sqrt{\frac{B_\beta}{\alpha_{ij} \alpha_{ij}}},
\end{align}
where $\alpha_{ij} = \partial \overline{u}_j / \partial x_i$,
$\beta_{ij} = \Delta^2 \alpha_{mi} \alpha_{mj}$, and $B_\beta =
\beta_{11} \beta_{22} - \beta_{12}^2 + \beta_{11}\beta_{33} -
\beta_{13}^2 + \beta_{22} \beta_{33} - \beta_{33}^2$. The model
constant $C_{\mathrm{vre}} \approx 0.07$.

For both the Vreman model and the DSM, the term $\tau_{kk}\delta_{ij}$
is neglected on the grounds that it is small compared to the
thermodynamic pressure.  The SGS model for the heat flux in the energy
equation is given by the eddy diffusivity model $q_i =
-\overline{\rho} \nu_t/\mathrm{Pr}_t \partial \widetilde{T}/\partial
x_i$, where $T$ is the temperature, and Pr$_t$ is the turbulent
Prandtl number which is set equal to 0.9.  The equation of state for
the fluid is a calorically perfect gas $\overline{p} = \overline{\rho}
R \widetilde{T}$. Additional details about the computational approach can
be found in \citet{Fu2022}.

\subsection{Wall Model}

\begin{figure}
\centering
\includegraphics[width=.6\textwidth]{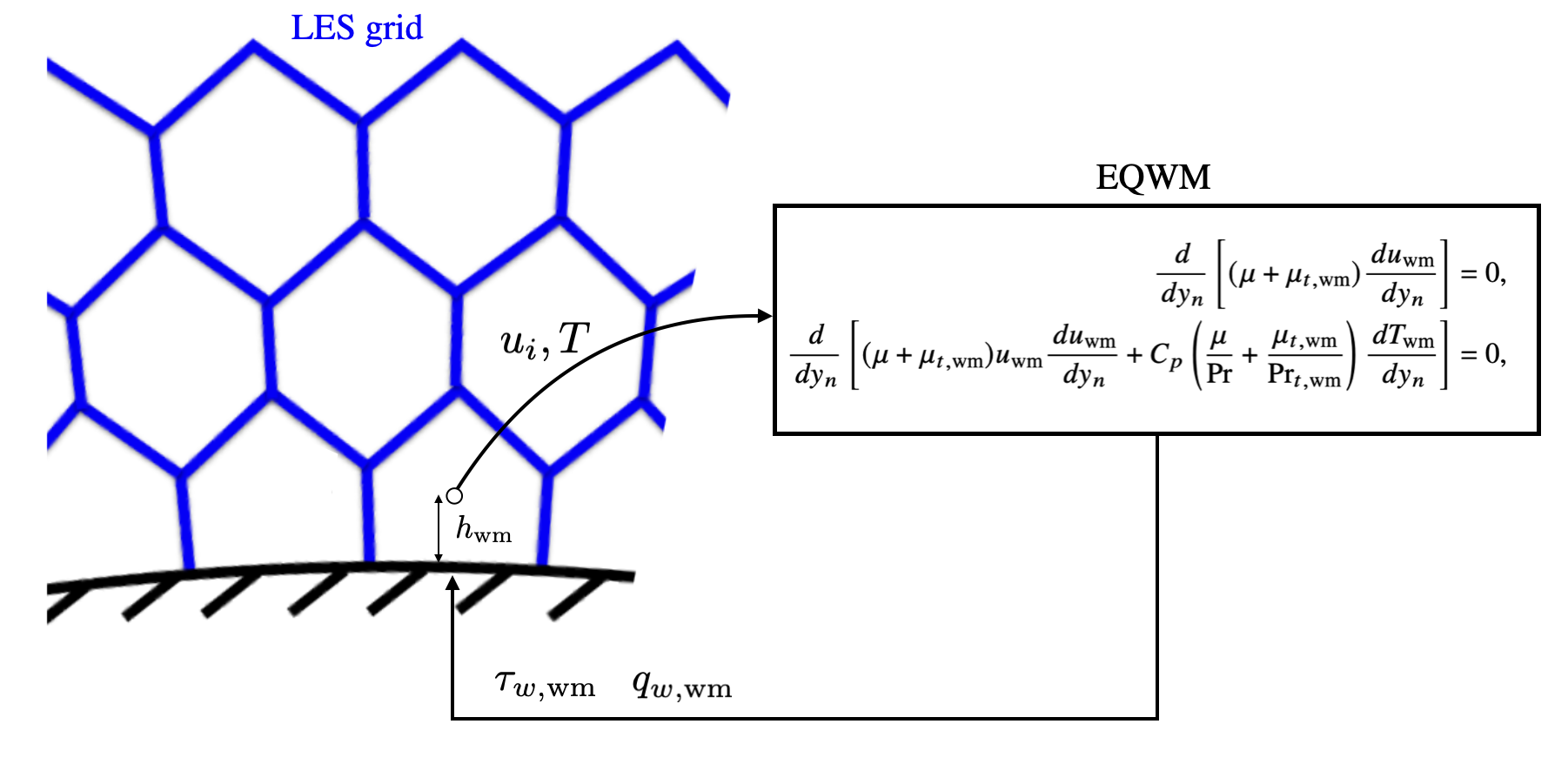}
\caption{Schematic of the equilibrium wall model.}
\label{wm-channel}
\end{figure}

We use a compressible equilibrium wall-stress model (EQWM) to model
the small-scale flow motions near the wall.  Figure~\ref{wm-channel}
shows a schematic of the wall modeling approach.
The form of the wall model is given by two ordinary differential
equations derived from the conservation of momentum and energy by
neglecting the wall-parallel fluxes~\cite{larsson2010wallmodeling},
\begin{subequations}
\label{eq:wallmodel}
\begin{align}
    \frac{\,d}{\,d y_n}\left[(\mu+\mu_{t,\text{wm}})\frac{\,d u_{\text{wm}}}{\,d y_n}\right] &= 0,\\
    \frac{\,d}{\,d y_n}\left[(\mu + \mu_{t,\text{wm}}) u_{\text{wm}} \frac{\,d u_{\text{wm}}}{\,d y_n} + C_p \left(\frac{\mu}{\text{Pr}} + \frac{\mu_{t,\text{wm}}}{\text{Pr}_{t,\text{wm}}}\right)\frac{\,d T_{\text{wm}}}{\,d y_n}\right] &= 0,
\end{align}
\end{subequations}
where $y_n$ is the wall-normal direction, $u_\mathrm{wm}$ is the
magnitude of the wall-parallel velocity for the wall model,
$T_\mathrm{wm}$ is the wall-model temperature, $\mu$ is the dynamic
viscosity, $\mu_{t,\mathrm{wm}}$ is the wall-model eddy viscosity, Pr and
Pr$_\mathrm{t,wm}$ are Prandtl number and turbulent Prandtl number,
respectively, and $C_p$ is the specific heat at constant pressure. The
eddy viscosity is taken from the mixing-length model as
\begin{align}
    \mu_{t,\text{wm}} = \kappa \overline{\rho} y_n \sqrt{\frac{\tau_{w,\text{wm}}}{\overline{\rho}}} \left[1 - \exp{\left(\frac{-y_n^+}{A^+}\right)}\right]^2,
\end{align}
where $\kappa =0.41$, Pr$_{t,wm}=0.9$, $A^+=17$, $\overline{\rho}$ is the
density, $\tau_{w,\mathrm{wm}}$ is the instantaneous wall shear stress, and
superscript + denotes quantities normalized by $\tau_w$ and $\mu$. At
$y_n = 0$, Eq. (\ref{eq:wallmodel}) is complemented with the non-slip
and adiabatic wall boundary conditions. At $y_n = h_{\text{wm}}$,
$u_{\text{wm}}$ and $T_{\text{wm}}$ are set equal to the instantaneous
values from the LES solution.  The matching location $h_{\text{wm}}$
is taken at the centroid of the first control volume attached to the
wall. Effectively, given an instantaneous wall-parallel velocity and
temperature at some height above the wall, Eq.~\ref{eq:wallmodel} is
solved for $u_{\text{wm}}$ and $T_{\text{wm}}$ in an overlapping layer
between $y_n = 0$ and $y_n = h_{\text{wm}}$. The solution from the
wall shear stress ($\tau_{w,\text{wm}}$) and the wall heat flux
($q_{w,\text{wm}}$) from the wall model are returned to the LES solver
as boundary conditions at the wall.

\subsection{Numerical Solver and Computational Resources}
\label{sec:solver}

The simulations are performed using the CPU version of the code charLES
from Cascade Technologies (Cadence). The validation of the algorithm
can be found in \citet{Fu2020}. The solver integrates the filtered
Navier-Stokes equations using a second-order accurate finite volume
formulation. The numerical discretization relies on a flux formulation
that is approximately entropy preserving in the inviscid limit,
thereby limiting the amount of numerical dissipation added into the
calculation. No special treatment was employed for the shock waves, and
the simulations in the present work are conducted without a shock
capturing scheme or sensor. The time integration is performed with a
third-order Runge-Kutta explicit method. The mesh generator is based
on a Voronoi hexagonal close packed point-seeding method which
automatically builds high-quality meshes for arbitrarily complex
geometries with minimal user input.

All simulations in this work are performed using high-performance
computing (HPC) resources provided by the MIT Supercloud. The MIT
Supercloud is a collaboration with MIT Lincoln Laboratory on a shared
facility that is optimized for streamlining open
research~\cite{reuther2018interactive}.  The system is designed to
support work that requires significant compute or memory
resources. The system has 480 nodes with Intel Xeon Platinum 8260
processors with 2 $\times$ 24 CPUs/node and 192GB of RAM per node.

\section{Computational Setup}
\label{sec:setup}

\begin{figure}
\centering
\includegraphics[width=\textwidth]{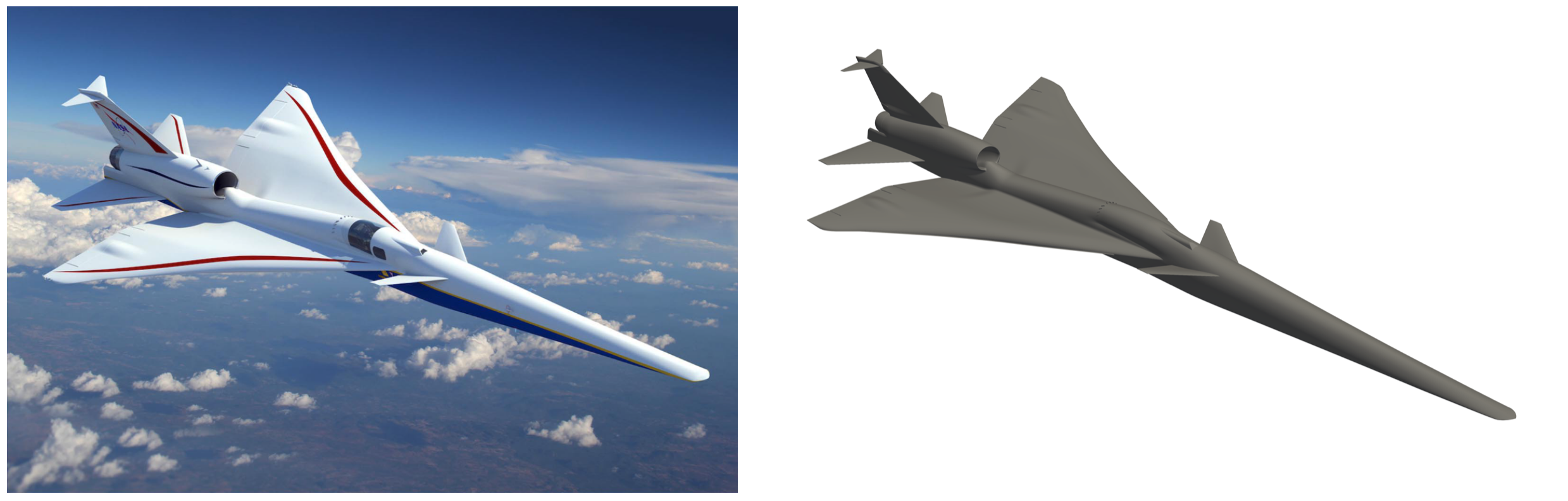}
\caption{The C608 configuration of the X-59 from
  \citet{park2022lowboom}.}
\label{x59-c608}
\end{figure}

\begin{figure}
\centering
\includegraphics[width=\textwidth]{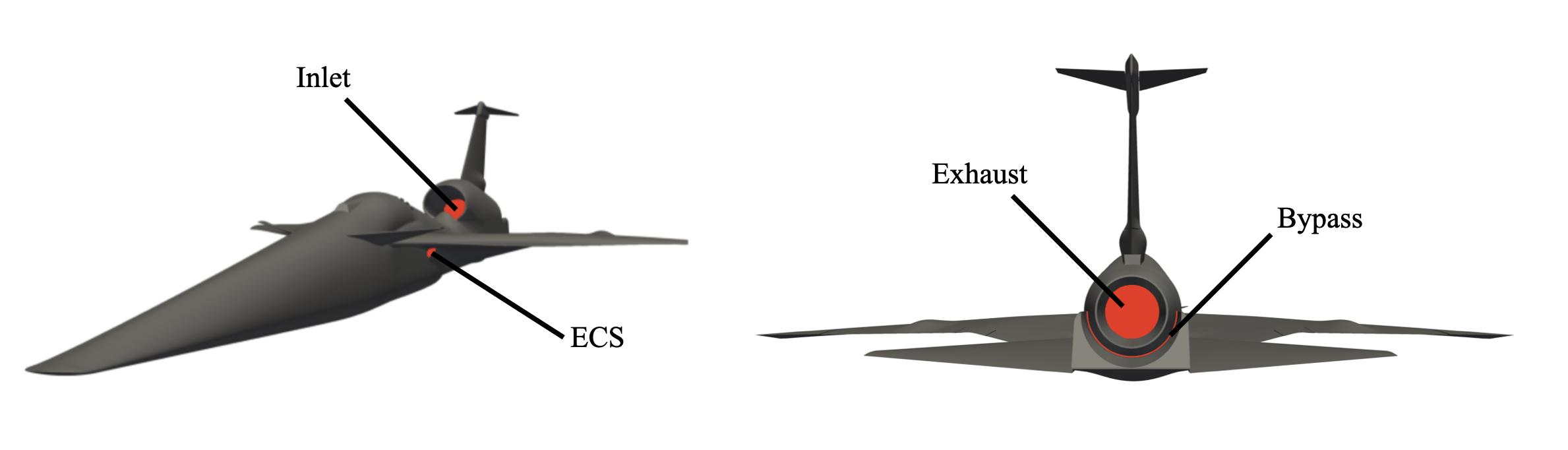}
\caption{C608 boundary conditions.}
\label{c608-bc}
\end{figure}

The simulation is performed using the C608 Low-Boom Flight
Demonstrator geometry provided by the AIAA Sonic Boom Prediction
workshop shown in Figure~\ref{x59-c608}. This geometry configuration
is an early iteration of the X-59 final design and is desirable for
comparing against RANS results using the same setup. Simulations are
performed at cruise conditions at Mach number M = 1.4 and Reynolds
number per inch of Re/in = 109,776. Freestream and boundary conditions
are specified in Table~\ref{tab:bcs}. The C608 has propulsion and
environmental control system (ECS) boundary conditions as shown in
Figure~\ref{c608-bc}. The engine bypass exhaust is a semicircular
region between the nozzle and aft deck, and the ECS inlet is located
in the wing root~\cite{park2022lowboom}.
\begin{table}
\caption{C608 freestream and boundary conditions}
\label{tab:bcs}
\begin{center}
\begin{tabular}{lc}
\hline\hline
Description & Condition\\ \hline
Freestream Mach number & 1.4 \\
Freestream temperature ($^\circ$R) & 389.9 \\
Altitude (ft) & 53,200 \\
Unit Reynolds number (per inch) & 109,776 \\
Ratio of engine nozzle plenum total pressure to freestream static pressure & 10.0 \\
Ratio of engine nozzle plenum total temperature to freestream static temperature & 7.0 \\
Ratio of engine bypass exhaust total pressure to freestream static pressure & 2.4 \\
Ratio of engine bypass exhaust total temperature to freestream static temperature & 2.0 \\
Ratio of engine fan face static pressure to freestream & 2.6 \\
Ratio of ECS inlet static pressure to freestream & 1.4\\
\hline\hline
\end{tabular}
\end{center}
\end{table}

\begin{figure}
  \centering
  \includegraphics[width=0.5\textwidth]{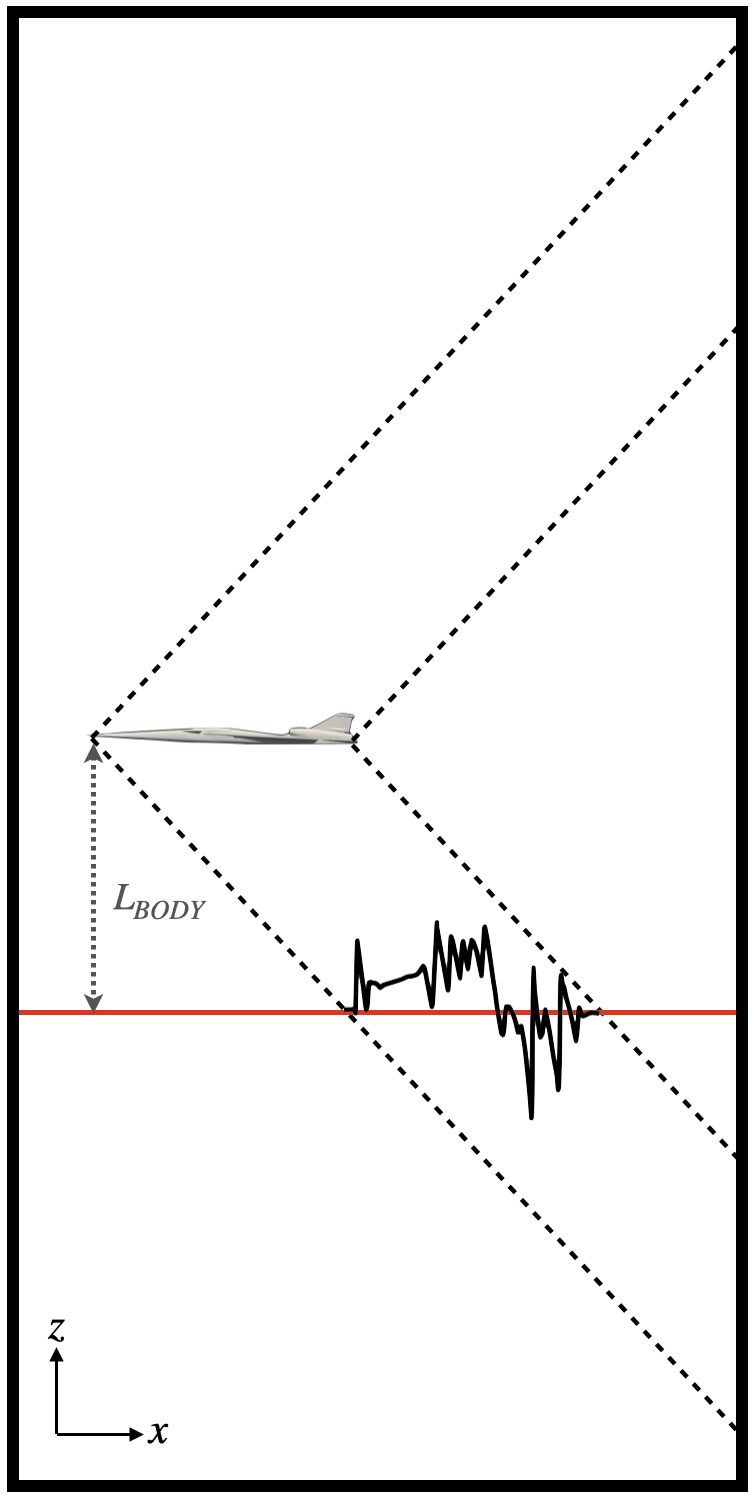}
  \caption{Computational domain for X-59 simulations.}
  \label{x59-domain}
\end{figure}

The size of the computational domain is $2.7L_{BODY} \times
2.7L_{BODY} \times 5.4L_{BODY}$ in the streamwise, 
spanwise, and vertical directions,
respectively, where $L_{BODY}$ is the streamwise
body-length of the model. Figure~\ref{x59-domain} shows a schematic of
the computational domain, where the red line denotes the pressure
probe placed at one-body length below directly under the aircraft.

\begin{figure}
\centering
    \begin{subfigure}[b]{0.45\textwidth}
        \centering
        \includegraphics[width=0.8\textwidth]{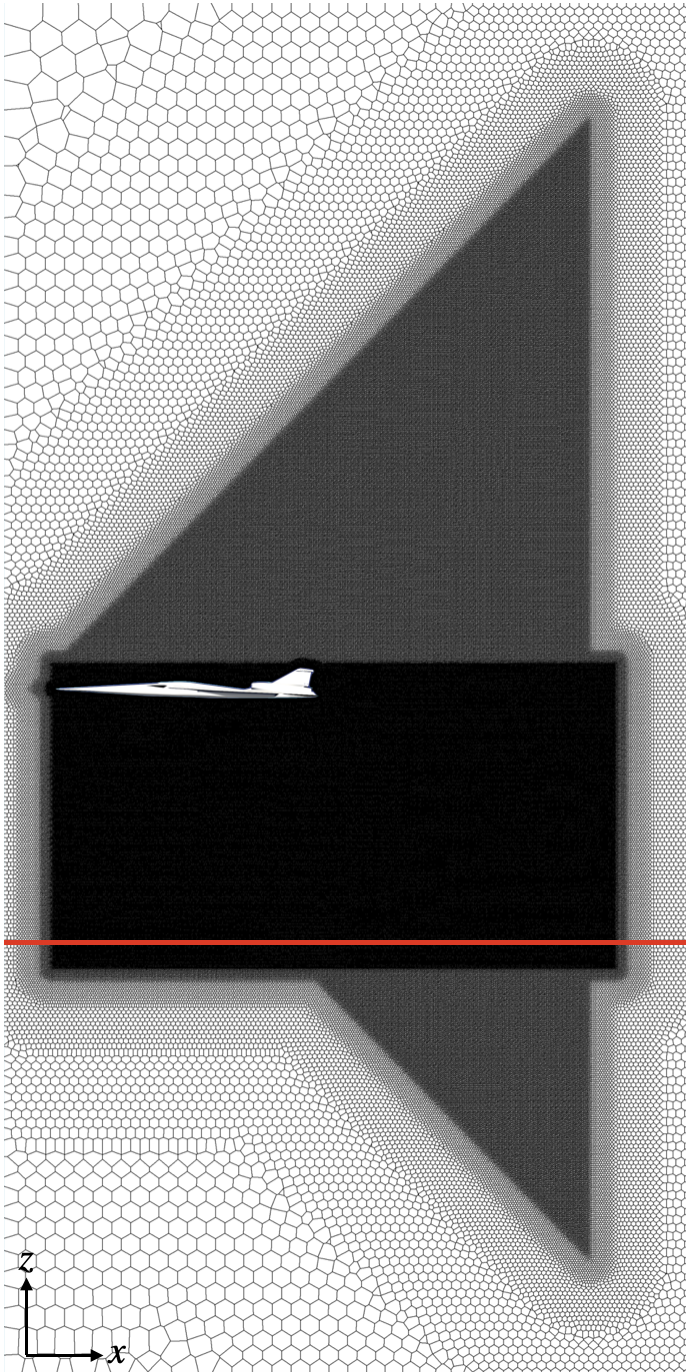}
        \caption{Side view of the WMLES grid topology across the
          $x$-$z$ plane. The red line indicates the location of the
          pressure probe.}
        \label{x59-grid-xz}
    \end{subfigure}
    \hfill
    \begin{subfigure}[b]{0.45\textwidth}
        \centering
        \includegraphics[width=0.8\textwidth]{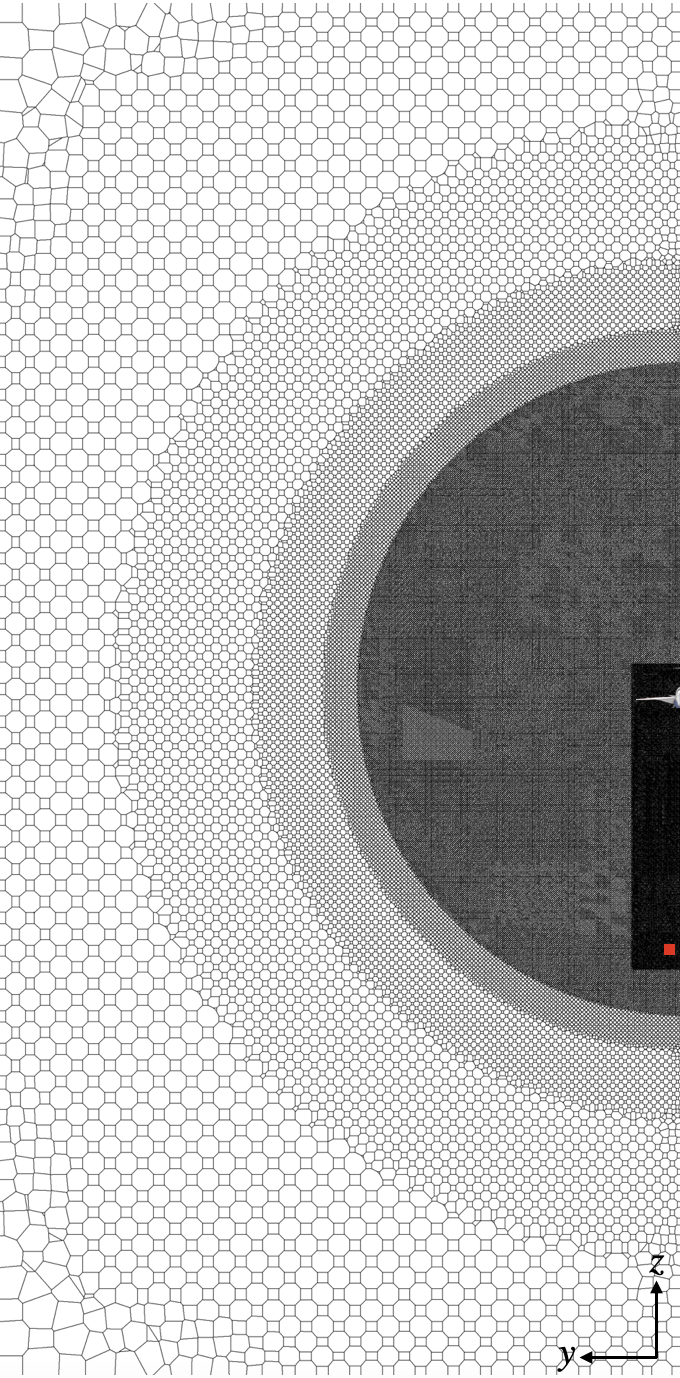}
        \caption{Front view of the WMLES grid topology across the
          $y$-$z$ plane. The red line indicates the location of the
          pressure probe.}
        \label{x59-grid-yz}
    \end{subfigure}
    \begin{subfigure}[b]{0.45\textwidth}
      \centering
      \includegraphics[width=\textwidth]{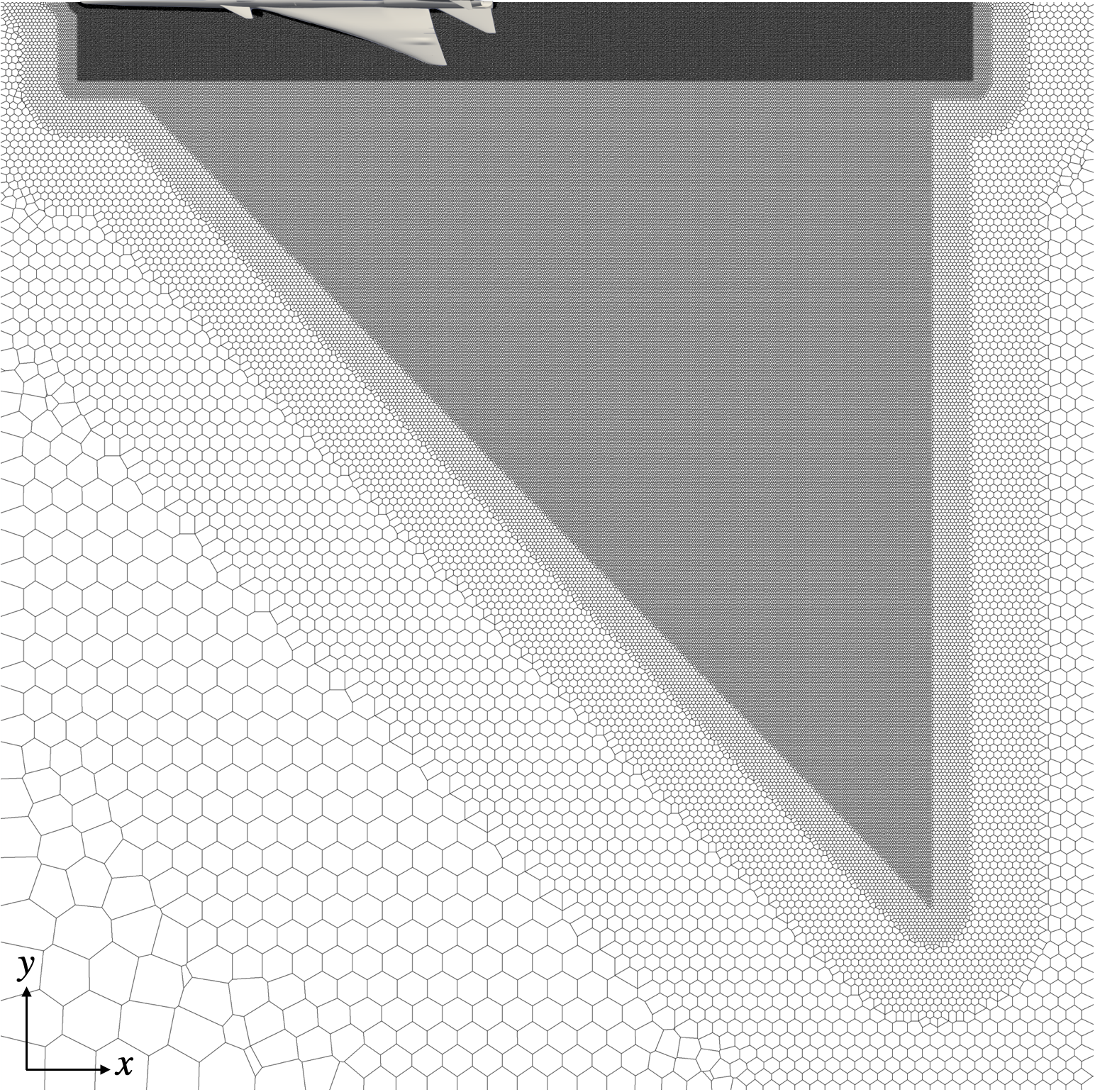}
      \caption{Top view of the WMLES grid topology across the $x$-$y$
        plane.}
      \label{x59-grid-xy}
    \end{subfigure}
    \caption{Grid topology for cases 3 and 6.}
\end{figure}

%
A total of six WMLES cases are conducted for three different grid
resolutions using both DSM and Vreman models.
In all cases, the domain is discretized using a Mach-aligned cone
of length $2.1L_{BODY}$ and whose vertex is located at $0.1L_{BODY}$ 
upstream the aircraft nose,
and an additional grid refinement in the probe-containing
plane where the pressure signatures are reported. 
This additional refined region extends from the aircraft nose to $2.2L_{BODY}$ in
the streamwise direction, 
to $0.25L_{BODY}$ in the spanwise direction (enclosing the
wing of the aircraft),
and to $1.5L_{BODY}$ below the aircraft.
The cases also
include enhanced refinement around the body of the aircraft geometry.
Table~\ref{tab:x59_cases} compiles the information for the six
cases, including the grid size in the Mach-aligned cone and the probe-containing 
refined plane.
\begin{table}
\caption{Summary of grid resolution and SGS model for X-59 flow cases
  conducted with WMLES}
\label{tab:x59_cases}
\begin{center}
\begin{tabular}{cccccc}
\hline\hline
\multicolumn{1}{c}{\multirow{2}{*}{Case}} & \multicolumn{2}{c}{Mach-aligned cone} & \multicolumn{2}{c}{Probe-containing plane} & \multicolumn{1}{c}{\multirow{2}{*}{SGS model}} \\ \cline{2-5}
\multicolumn{1}{c}{} & \multicolumn{1}{c}{in/cell} & \multicolumn{1}{c}{Points/$L_{BODY}$} & \multicolumn{1}{c}{in/cell} & \multicolumn{1}{c}{Points/$L_{BODY}$} & \multicolumn{1}{c}{} \\ \hline

1 & 20 & 54 & 10 & 108 & DSM\\ 
2 & 10 & 108 & 5 & 216 & DSM\\ 
3 & 5 & 216 & 2.5 & 432 & DSM\\ 
4 & 20 & 54 & 10 & 108 & Vreman\\ 
5 & 10 & 108 & 5 & 216 & Vreman\\ 
6 & 5 & 216 & 2.5 & 432 & Vreman\\ \hline\hline
\end{tabular}
\end{center}
\end{table}
%
As an example, Figure~\ref{x59-grid-xz} shows the domain
discretization for one of the WMLES cases in a slice of the $x$-$z$
plane. Figure~\ref{x59-grid-yz} shows a view of the domain from the
front of the aircraft at a slice halfway through the domain. The front
and side views show the refinement configuration with a red line to
denote the pressure probe. Figure~\ref{x59-grid-xy} shows a top view
of the domain discretization.

\section{Results}
\label{sec:results}

\subsection{Shock Waves}

\begin{figure}
\centering
    \begin{subfigure}[b]{0.45\textwidth}
        \centering
        \includegraphics[width=\textwidth]{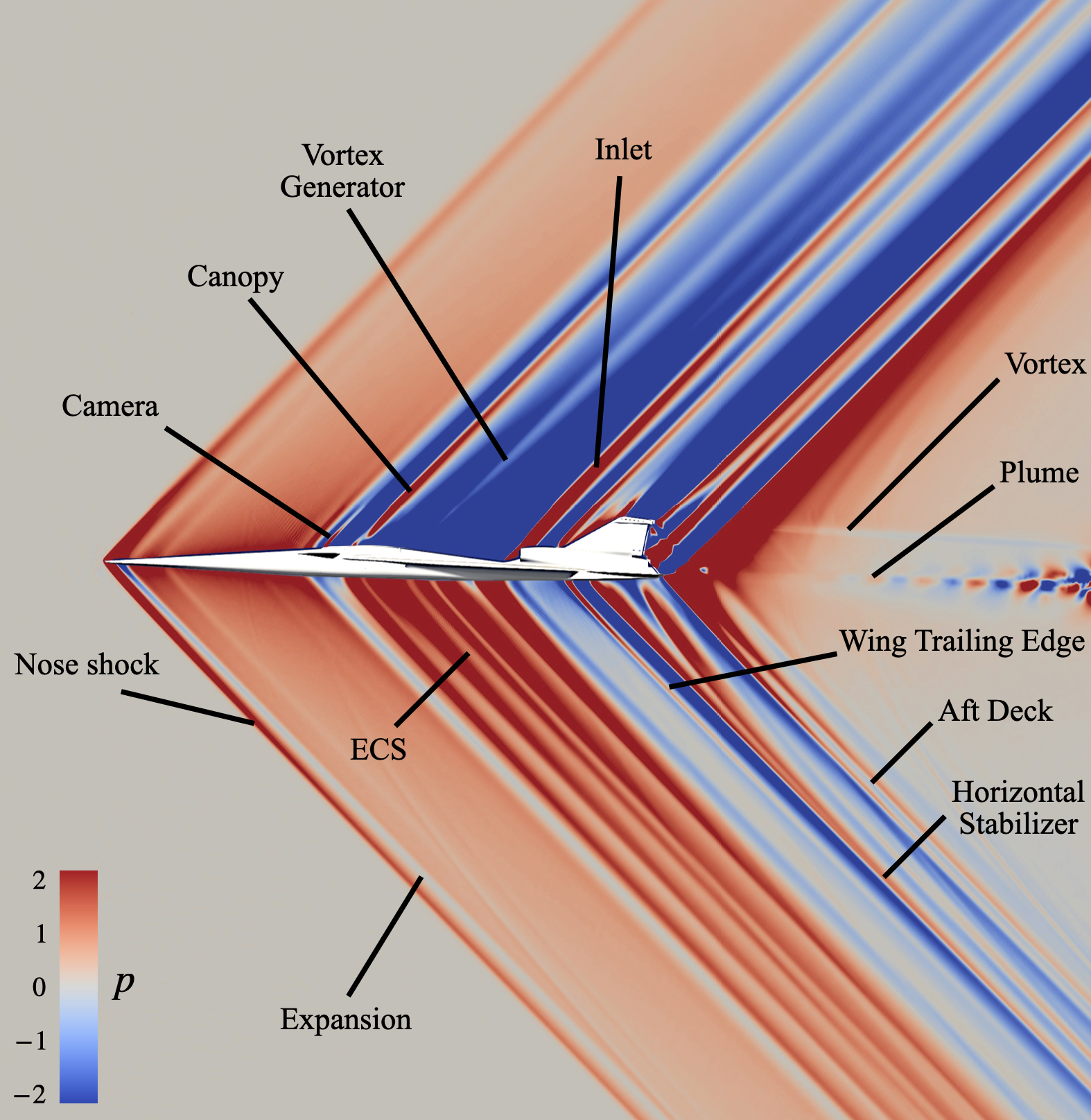}
        \caption{}
        \label{x59-shocks-side}
    \end{subfigure}
    \hfill
    \begin{subfigure}[b]{0.45\textwidth}
        \centering
        \includegraphics[width=\textwidth]{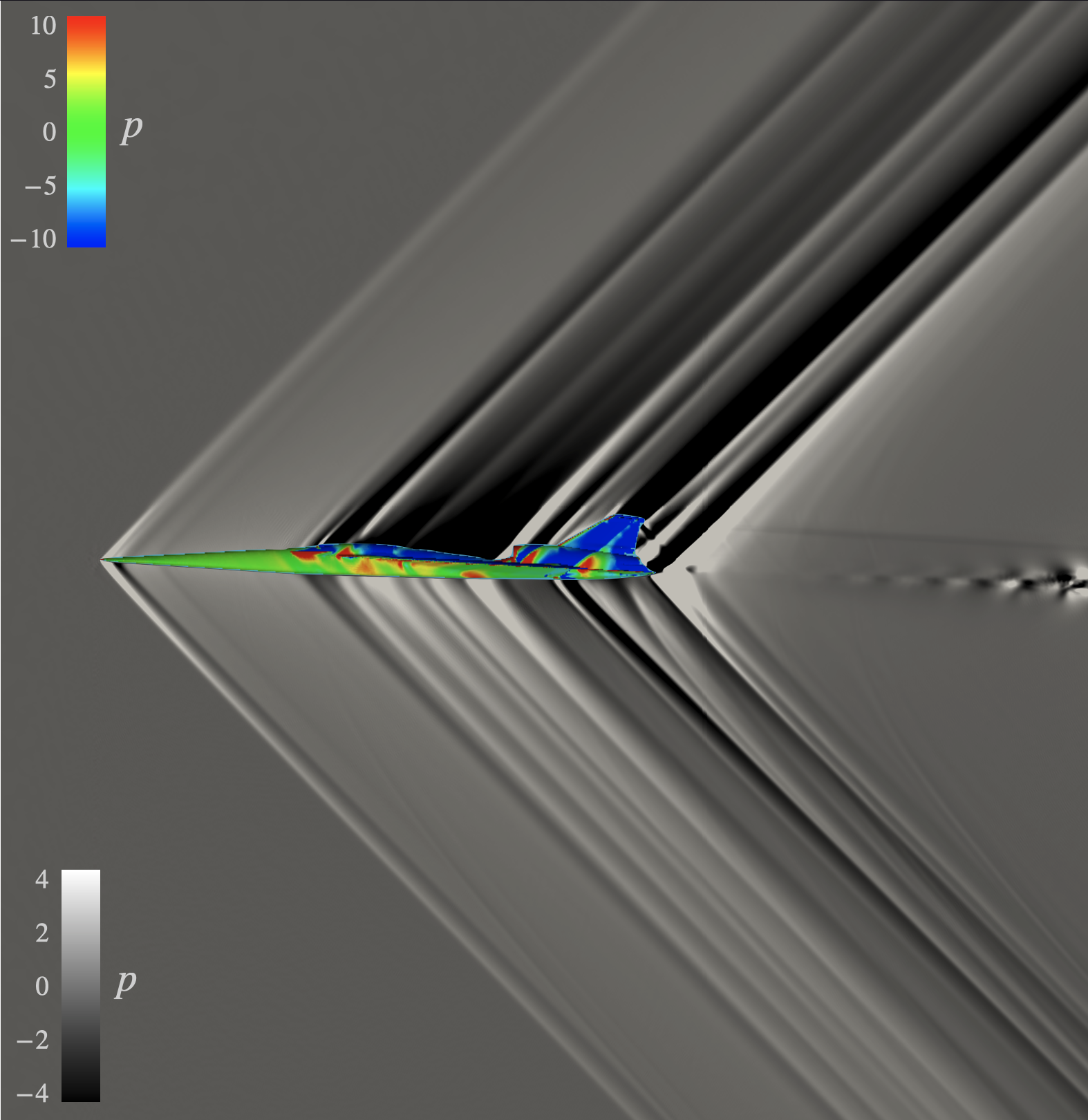}
        \caption{}
        \label{x59-schlieren-side}
    \end{subfigure}
    \caption{Pressure visualizations for case 3. (a) Pressure contours in the $y/L_{BODY}=0$ plane and (b) side view of the pressure distribution over the aircraft.}
\end{figure}

\begin{figure}
\centering
    \begin{subfigure}[b]{0.45\textwidth}
        \centering
        \includegraphics[width=\textwidth]{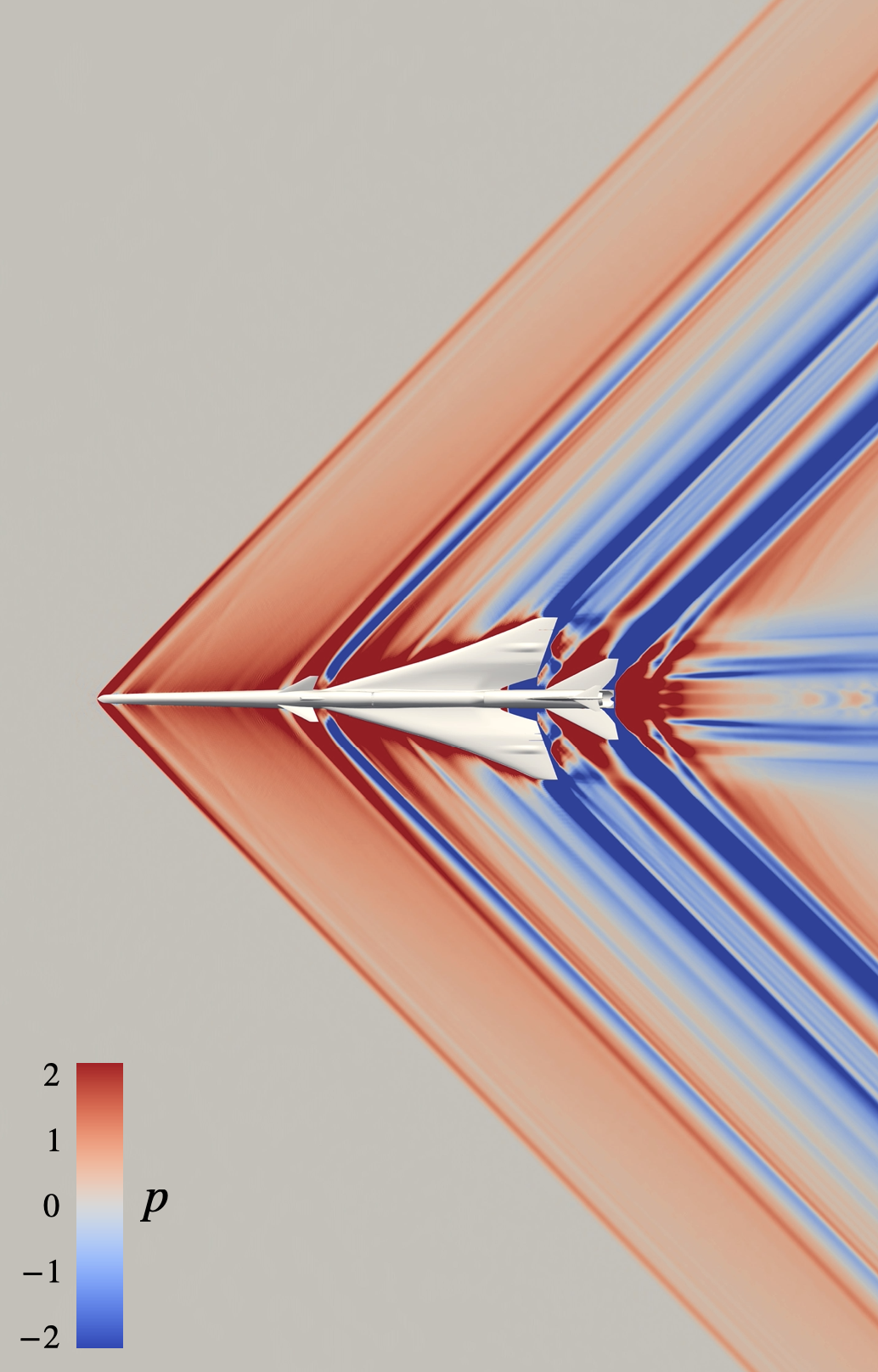}
        \caption{}
        \label{x59-shocks-top}
    \end{subfigure}
    \hfill
    \begin{subfigure}[b]{0.45\textwidth}
        \centering
        \includegraphics[width=\textwidth]{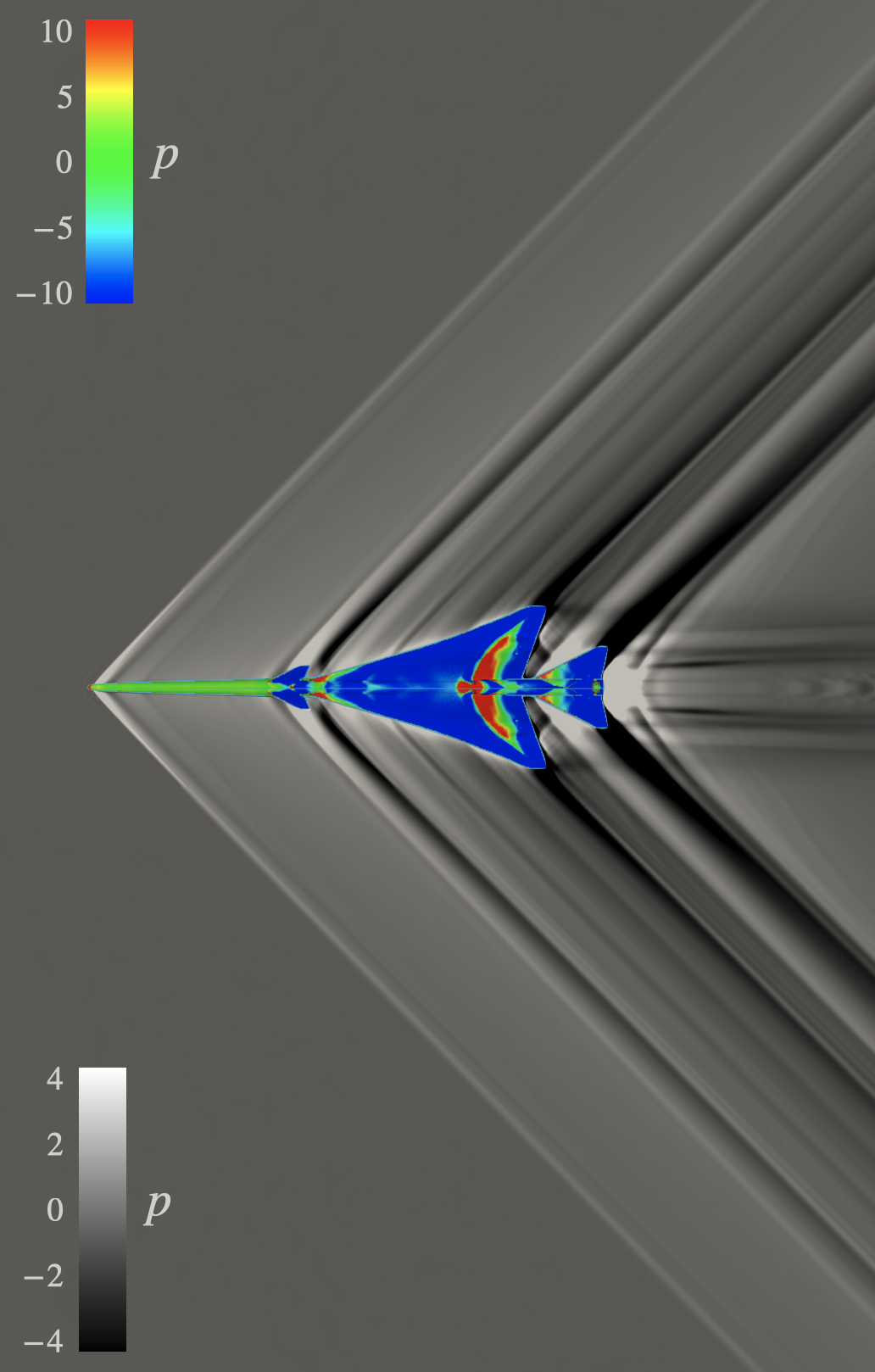}
        \caption{}
        \label{x59-schlieren-top}
    \end{subfigure}
    \caption{Pressure visualizations for case 3 (cont.). (a) Pressure contours in the $z/L_{BODY}=0$ plane and (b) top view of the pressure distribution over the aircraft.}
\end{figure}

First, we discuss the results for the locations and intensities of
shock waves observed from both the side and top views, depicted in
Figure~\ref{x59-shocks-side} and Figure~\ref{x59-shocks-top},
respectively. We label the shocks that arise due to the aircraft's
geometry and the boundary conditions.  Additionally, we provide
computational flow visualizations of the side and top views in
Figure~\ref{x59-schlieren-side} and Figure~\ref{x59-schlieren-top},
respectively. These visualizations are generated to capture the
contours in the relative pressure, denoted as $p$, which is defined
as the percent difference deviation from the freestream pressure
\begin{align}
    p = \frac{\Delta p}{p_\infty} \times 100,
\end{align}
where $\Delta p = p_s - p_\infty$ where $p_s$ is the pressure measured by the probe and
\begin{align}
    p_\infty = \frac{\rho_\infty}{\gamma} \left(\frac{U_\infty}{\text{M}}\right)^2.
\end{align}
where the subscript $\infty$ denotes freestream quantity, $\gamma =
1.4$, and M = 1.4. The pressure visualizations show the surface
pressure distribution on the aircraft, also quantified using $p$
above. The nose creates a shock and expansion pair, followed by a
smooth compression. There is a rapid succession of shocks and
expansions generated by the wing leading edge, vortex generators, and
ECS inlet. The wing trailing edge, horizontal stabilizer, aft deck,
tail, plume, and vortices result in additional shocks and expansions
along with other complex interactions. The surface pressure
distribution indicates that the inlet spillage creates an area of high
pressure over the wing. These results are consistent with those
reported through the AIAA Sonic Boom Prediction
Workshop~\cite{park2022lowboom}.

\subsection{Pressure Signatures}

The WMLES pressure signatures at one-body length away from the
aircraft are compared with RANS results from \citet{kirz2022dlrtau}
for a grid resolution of 5.74 in/cell.  We define a transformed
x-coordinate given by
\begin{align}
    X_N = x - \frac{L_{BODY}}{\tan(\mu_M)},
\end{align}
where the distance from the nose in the freestream direction $x$ is
normalized by the Mach angle $\mu_M = \sin^{-1}(1/\text{M})$ using the
radial distance to the model, $L_{BODY}$ in this case. This
transformation is common in previous sonic boom
simulations~\cite{kirz2022dlrtau}.

\begin{figure}
    \begin{subfigure}[t]{0.45\textwidth}
        \centering
        \includegraphics[width=\textwidth]{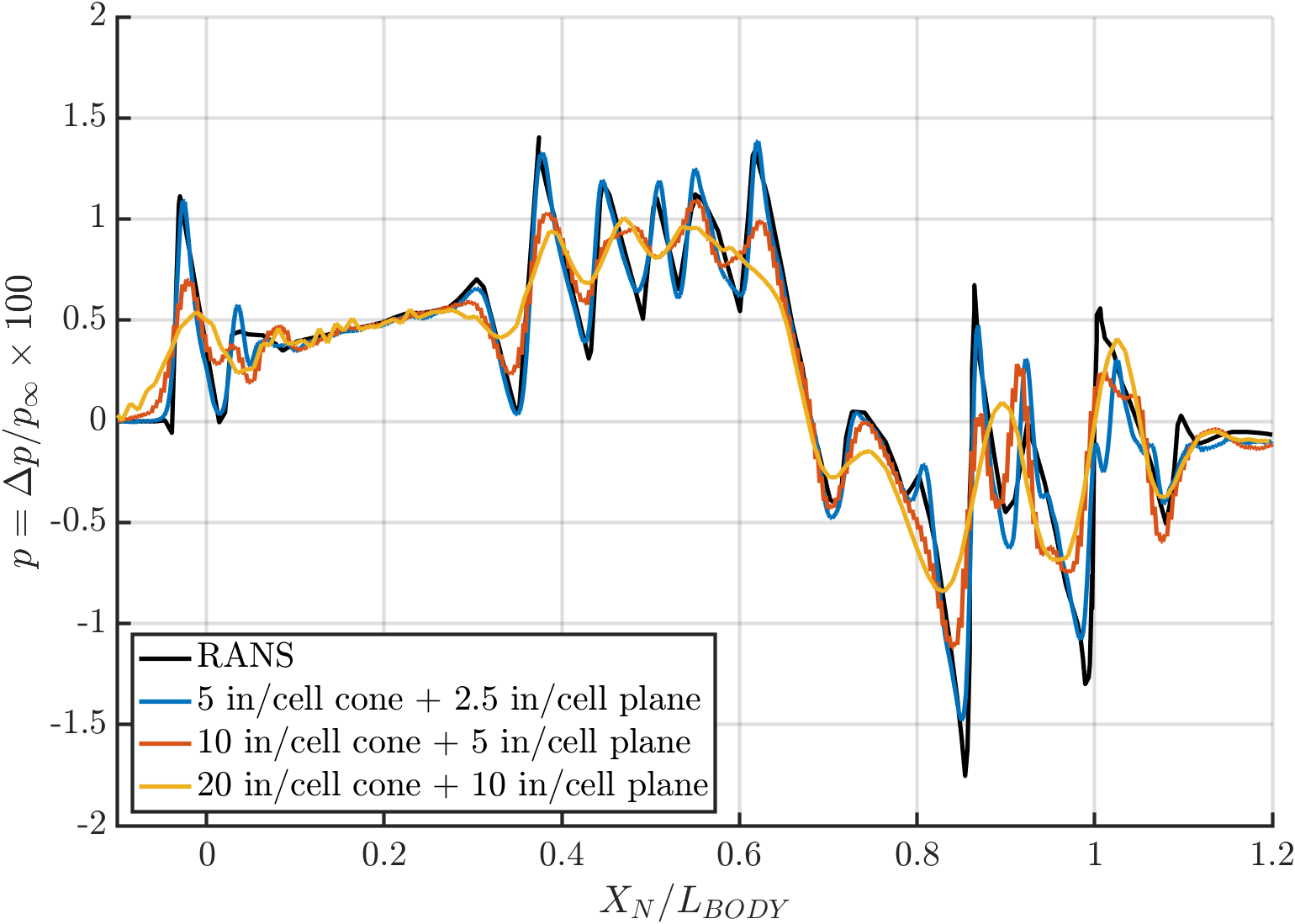}
        \caption{Pressure signature for DSM and different resolutions
          (cases 1-3) as a function of $X_N/L_{BODY}$.}
        \label{x59-dsm-signature}
    \end{subfigure}
    \hfill
    \begin{subfigure}[t]{0.45\textwidth}
        \centering
        \includegraphics[width=\textwidth]{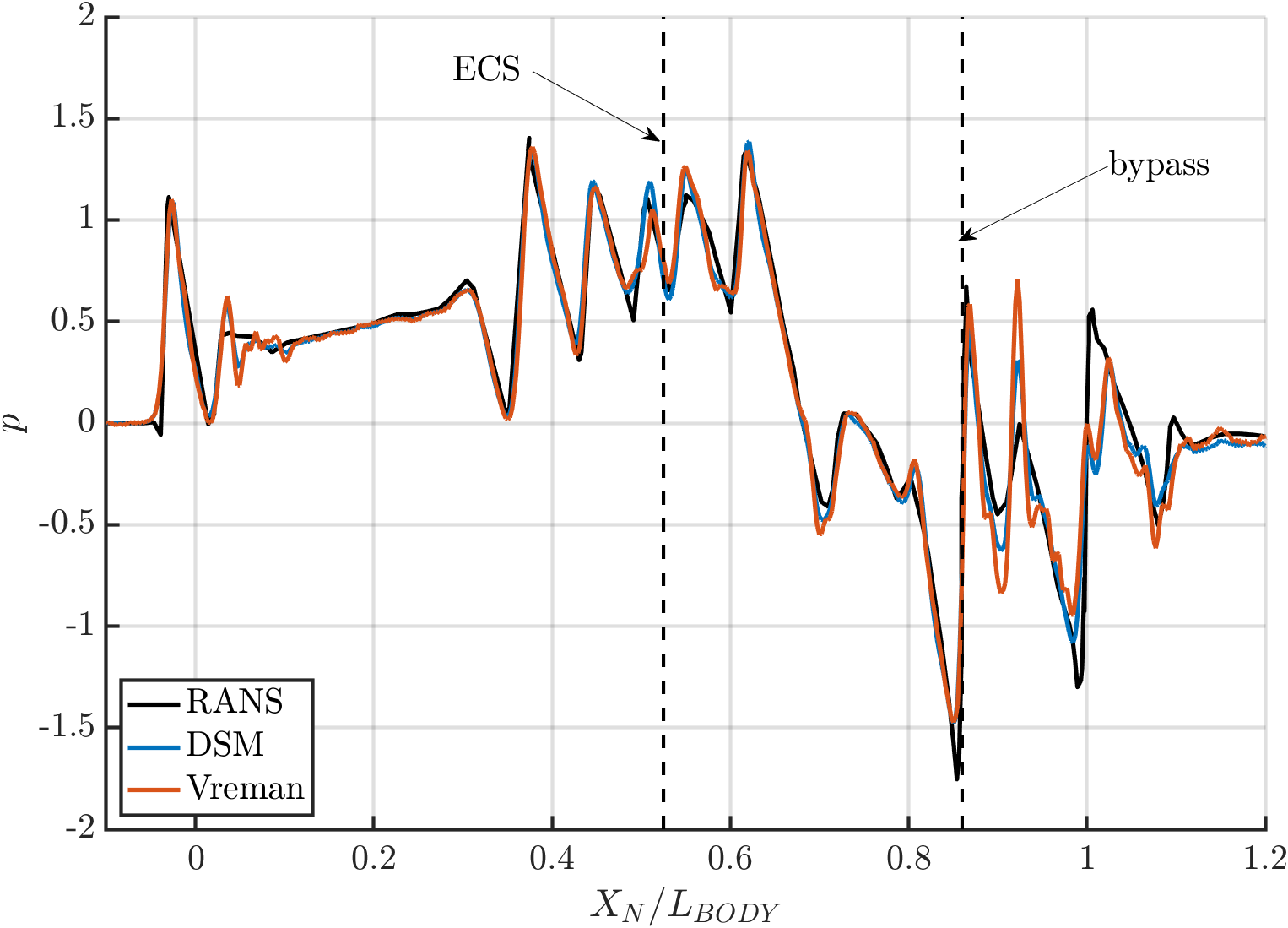}
        \caption{Pressure signature for DSM and Vreman for the finest
          grid resolution considered (cases 3 and 6) as a function of
          $X_N/L_{BODY}$.}
        \label{x59-vreman-signature}
    \end{subfigure}
    \caption{Near-field pressure signatures.}
\end{figure}

Figure~\ref{x59-dsm-signature} shows the prediction of the pressure
signature by WMLES with the DSM for the different grid refinement
configurations outlined in Table~\ref{tab:x59_cases}. As the grid is
refined, the prediction provided by WMLES approaches the RANS results
for the near-field pressure signature. Results computed using the
Vreman SGS model were comparable and are omitted for the sake of
brevity.  Figure~\ref{x59-vreman-signature} shows the finest
resolution case comparison with both SGS models.

The WMLES solutions show reasonably accurate agreement with the RANS
results for $X_N/L_{BODY} < 0.9$ for both SGS models, particularly at
the finest grid resolution. This is also the case in the vicinity of
the ECS and bypass, located approximately at $X_N/L_{BODY} \approx
0.45$ and $X_N/L_{BODY} \approx 0.85$, respectively. However, we
observe that WMLES results do not match the RANS solution as
effectively in the downstream region and after the aircraft. This
disparity is present for both SGS models and persists for the finest
grid resolution considered. Although the specific cause of this
disparity remains unclear, it is likely attributed to the intricate
interactions and rapid sequence of shocks and expansions occurring
near various components, including the wing trailing edge, horizontal
stabilator, aft deck, T-tail, engine plume, and tip
vortices~\cite{park2022lowboom}.

\begin{figure}
\centering
\includegraphics[width=0.4\textwidth]{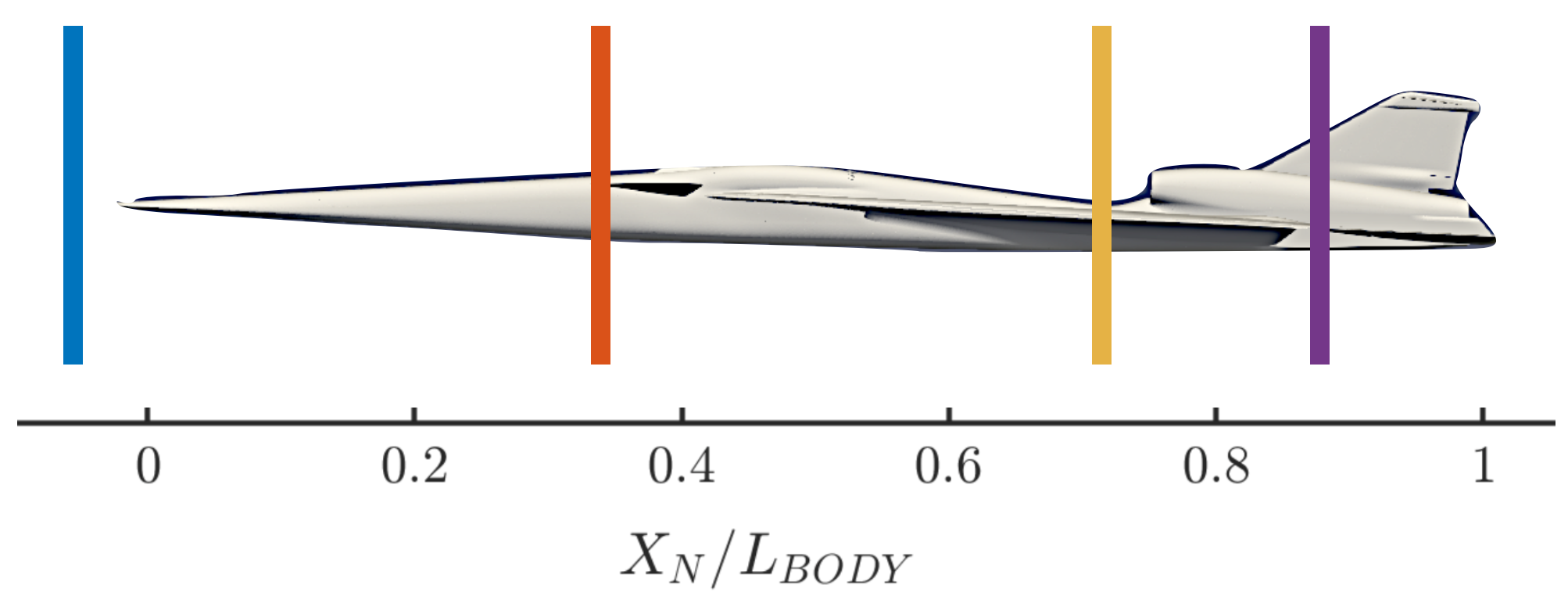}
\caption{Schematic of aircraft with delineations at specified locations.}
\label{x59-slice}
\end{figure}

\begin{figure}
\centering
    \begin{subfigure}[t]{0.45\textwidth}
        \centering
        \includegraphics[width=\textwidth]{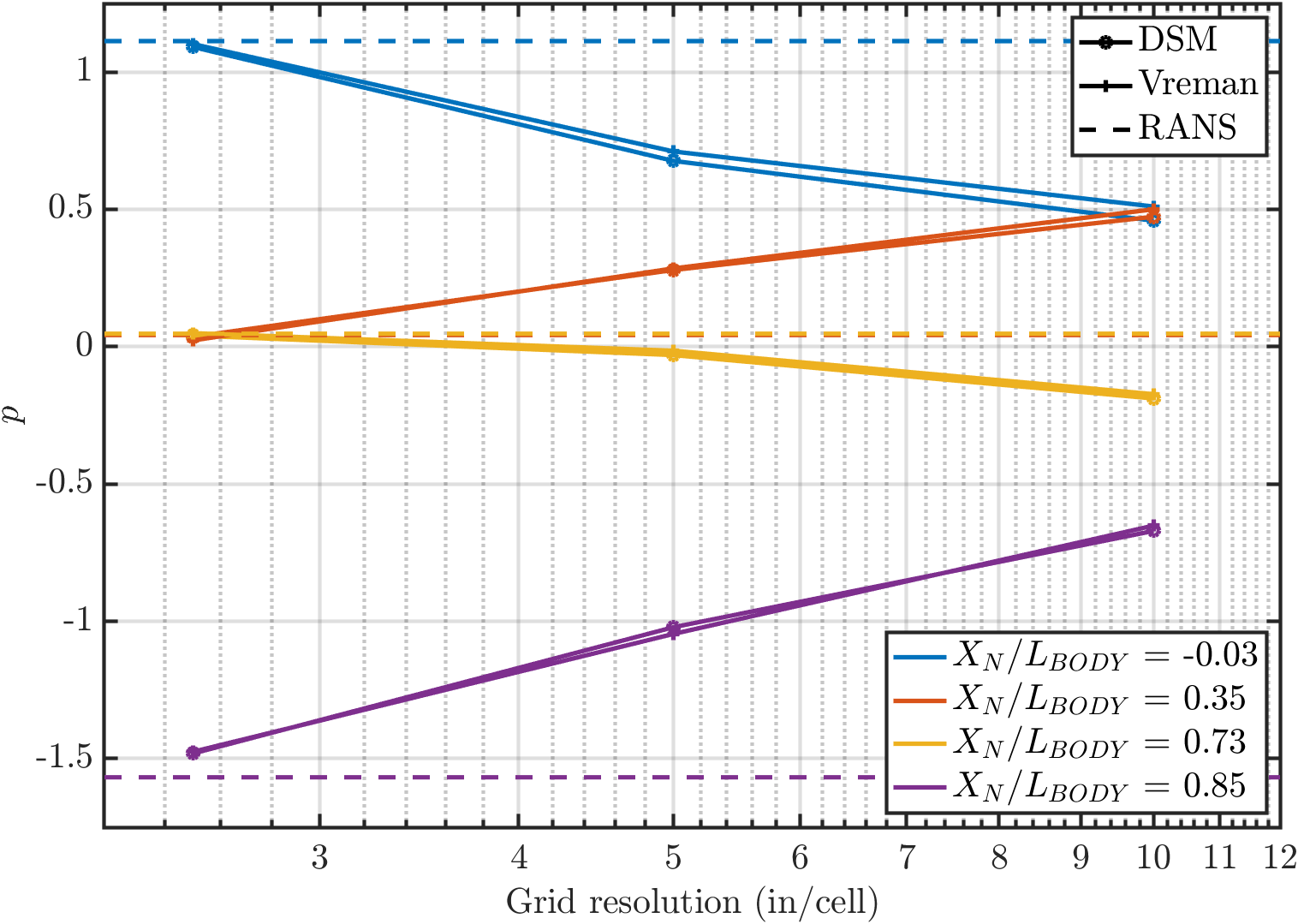}
        \caption{Pressure signature as a function of grid resolution.}
        \label{x59-convergence}
    \end{subfigure}
    \hfill
    \begin{subfigure}[t]{0.45\textwidth}
        \centering
        \includegraphics[width=\textwidth]{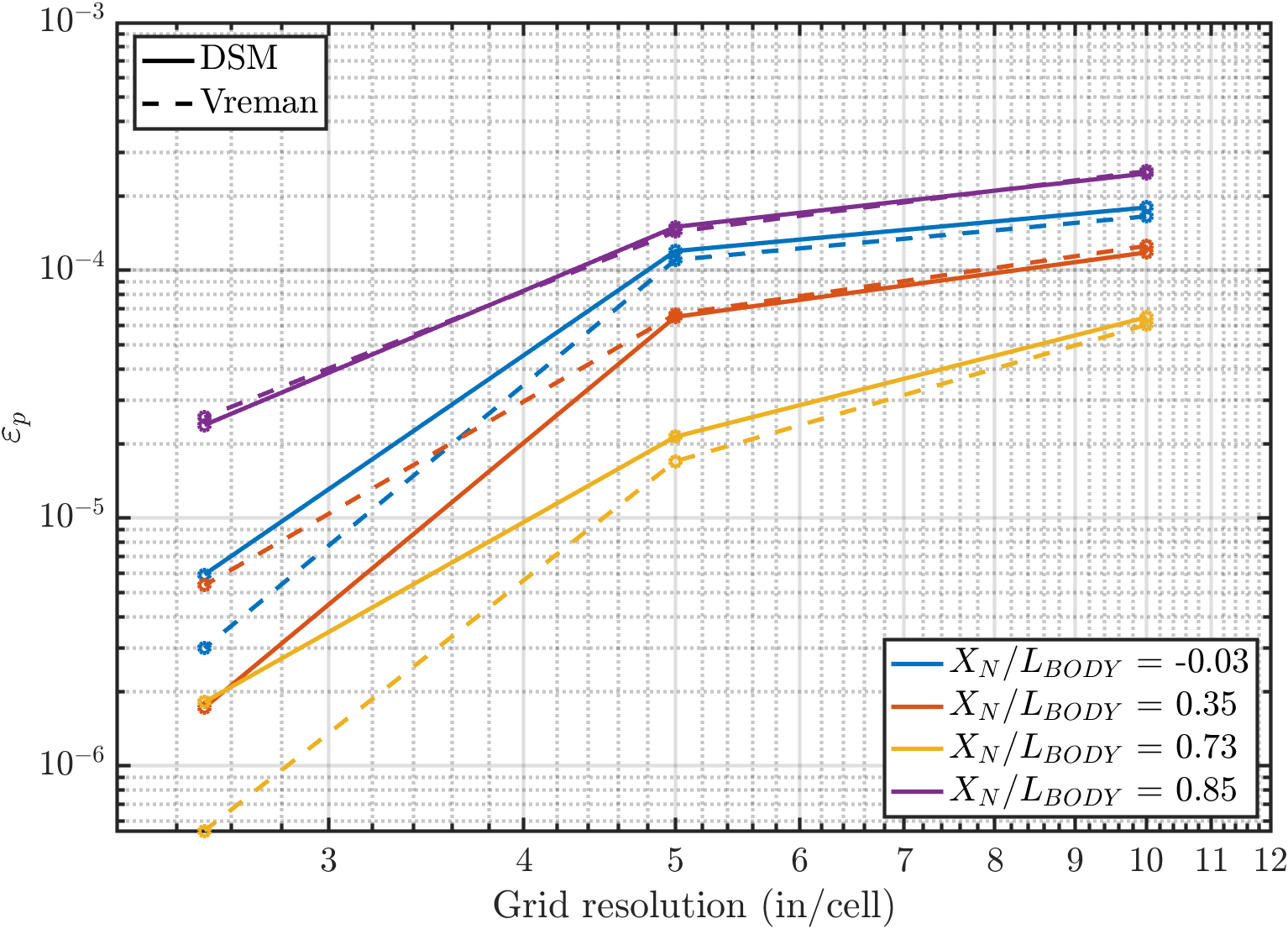}
        \caption{Relative pressure signature difference,
          $\varepsilon_p$, as a function of grid resolution.}
        \label{x59-error}
    \end{subfigure}
    \caption{Pressure and $\varepsilon_p$ for near-field pressure
      signature prediction.}
\end{figure}

Figure~\ref{x59-convergence} contains the results for the pressure
signature as the grid is refined. The $X_N/L_{BODY}$ locations of
interest are shown in the schematic in Figure~\ref{x59-slice} for
reference to give insight into which areas of the geometry are
impactful at the location of the pressure probe. The RANS solution is
also included for reference as dashed lines, which is not dependent
on the grid resolution.
The relative difference in the pressure signature between the RANS and
WMLES is computed as
\begin{align}
    \varepsilon_p = \frac{p_{\text{WMLES}} - p_{\text{RANS}}}{p_\infty},
\end{align}
where the subscripts denote either the WMLES or RANS solution. The
results for $\varepsilon_p$, shown in Figure~\ref{x59-error}, reveal
that the difference between WMLES and RANS solution decreases with
grid refinement for all locations considered and both SGS models.

\subsection{Computational Cost}
\label{sec:cost}

Cost-efficiency is part of the motivation for using WMLES for
turbulence modeling.  Table~\ref{tab:cost} outlines the calculations
for the computational cost for different grid refinement
configurations for the WMLES cases with the Vreman SGS model. This
estimate is based on five flow passes through the domain to ensure
convergence in the solution of the pressure signature propagated to
one-body length below the aircraft. The number of characteristic steps
in the simulation is determined using $\Delta t$ estimated from the
Courant number. The computational time in core-s is then converted to
CPU-hr. 
\begin{table}[H]
\caption{Computational cost estimates for WMLES with Vreman SGS model.}
\label{tab:cost}
\begin{center}
\begin{tabular}{c|c|c|c}\hline\hline
 & \multicolumn{1}{c|}{\begin{tabular}[c]{@{}c@{}}20 in/cell cone +\\ 10 in/cell plane\end{tabular}} & \multicolumn{1}{c|}{\begin{tabular}[c]{@{}c@{}}10 in/cell cone +\\ 5 in/cell plane\end{tabular}} & \multicolumn{1}{c}{\begin{tabular}[c]{@{}c@{}}5 in/cell cone +\\ 2.5 in/cell plane\end{tabular}} \\\hline
CV (M) & 3 & 18 & 135 \\\hline
$\Delta t$ (s) & 0.15 & 0.075 & 0.0375  \\\hline
Characteristic steps & 25,000 & 50,000 & 100,000 \\\hline
\begin{tabular}[c]{@{}l@{}}Normalized time\\ (core-s/CV/step)\end{tabular} & \multicolumn{3}{c}{12}\\\hline
Total nodes & \multicolumn{2}{c|}{5} & 10 \\\hline
Total cores & \multicolumn{2}{c|}{240} & 480 \\\hline
\textbf{CPU-hr} & 250 & 3000 & 45,000 \\\hline\hline
\end{tabular}
\end{center}
\end{table}

Comparing the costs between WMLES and RANS simulations is not a
straightforward task. In this regard, we refer to the cost reported by
\citet{spurlock2022cartesianmesh} as a
reference. \citet{spurlock2022cartesianmesh} performed RANS
simulations for the AIAA Sonic Boom Prediction Workshop on the X-59
aircraft using the NASA Ames Electra cluster. Their setup involved one
Skylake node equipped with two Intel Xeon Gold 6148 processors,
utilizing a total of 40 cores for each run. The total computational
time for their finest adapted grid, consisting of 29.6 million cells,
was reported to be 1431 CPU-hours. As an example, the WMLES simulation
using a grid configuration of 10 inches per cell for the cone and 5
inches per cell for the plane contains 18 million control volumes,
which is comparable to the RANS simulation with 29.6 million control
volumes. The estimated total computational time for WMLES using this
specific solver is approximately 3000 CPU-hours. This indicates that,
for a comparable grid resolution, WMLES is computationally more
demanding than RANS but still exhibits a similar cost.  It is
important to note that improvements in computational cost can be
achieved by optimizing the solver, utilizing faster CPUs, or employing
GPUs, which can potentially lead to 7 to 10 times faster turnaround
times. Nevertheless, our cost estimates highlight the motivation for
future efforts in developing SGS models that offer higher accuracy for
relatively coarser grids. Such advancements would directly result in
reduced computational costs for WMLES solutions.
 
\section{Conclusions}
\label{sec:conclusion}

Current RANS methodologies have shown reasonable performance in
predicting shock waves in regions far from their interactions with
boundary layers. However, a practical RANS model capable of accurately
addressing the broad range of flow regimes relevant to aerodynamic
vehicles has yet to be developed. In recent years, WMLES has emerged
as a tool for aerospace applications, offering a higher level of
generality compared to RANS approaches. Nevertheless, the application
of WMLES for predicting realistic supersonic aircraft has received
less attention. In this work, we evaluated the performance of WMLES
for realistic supersonic aircraft.

WMLES was utilized to analyze the experimental aircraft X-59 QueSST developed by Lockheed Martin at Skunk Works for
NASA's Low-Boom Flight Demonstrator project. The simulations employed
the charLES solver and aimed to assess the ability of WMLES to predict
near-field noise levels under cruise conditions. The study
considered two SGS models, the dynamic Smagorinsky
model and the Vreman model, combined with an equilibrium wall model. The
grid resolutions ranged from 20 in/cell to 5 in/cell in different areas of interest in the domain. The results
were compared with previous numerical studies based on the
RANS equations.

Our findings demonstrated that WMLES produced near-field pressure
predictions that were similar to those of RANS simulations for the finest grid resolutions
investigated at a
comparable computational cost. This was the case for both
SGS models. Some discrepancies were observed between the
WMLES and RANS predictions downstream the aircraft. These
differences persisted for the finest grid refinement considered,
suggesting that they might be attributed to the intricate interactions
of shock waves and expansion waves at the trailing edge. Note that the
refinement for the finest grid resolution was limited to a rectangular
region below the aircraft. As such, the interactions of shock waves
and expansion waves outside the refined area might be still severely
underresolved.

Here, our focus has been on WMLES with traditional SGS models along
with comparisons with RANS predictions. Future simulations of the X-59
Low-Boom Flight Demonstrator will follow the wind tunnel experiments
conducted NASA Glenn 8- by 6-Foot Supersonic Wind
Tunnel~\cite{Durston2023, Winski2023}.  Our long-term objective is to
develop a SGS model suitable for WMLES that offers accurate
predictions across various flow regimes, encompassing both subsonic
and supersonic flows. The findings obtained from the present study
will guide the advancements needed to expand our current
machine-learning based SGS modeling efforts~\cite{Lozano_brief_2020,
  Ling2022, Arranz2023, Lozano2023} in the presence of shock waves.

\section*{Funding Sources}

This material is based upon work supported by the U.S. Department of
Energy, Office of Science, Office of Advanced Scientific Computing
Research, Department of Energy Computational Science Graduate
Fellowship under Award Number DE-SC0023112. This paper was prepared as
an account of work sponsored by an agency of the United States
Government. Neither the United States Government nor any agency
thereof, nor any of their employees, makes any warranty, express or
implied, or assumes any legal liability or responsibility for the
accuracy, completeness, or usefulness of any information, apparatus,
product, or process disclosed, or represents that its use would not
infringe privately owned rights. Reference herein to any specific
commercial product, process, or service by trade name, trademark,
manufacturer, or otherwise does not necessarily constitute or imply
its endorsement, recommendation, or favoring by the United States
Government or any agency thereof. The views and opinions of authors
expressed herein do not necessarily state or reflect those of the
United States Government or any agency thereof.

\section*{Acknowledgments}
The authors acknowledge the MIT Supercloud and Lincoln Laboratory
Supercomputing Center for providing HPC resources that have
contributed to the research results reported within this paper.

\bibliography{main}

\end{document}